# On the Factors Influencing the Choices of Weekly Telecommuting Frequencies of Post-secondary Students in Toronto

Khandker Nurul Habib, Ph.D., PEng
Percy Edward Hart Professor in Civil & Mineral Engineering,
Associate Director, Data Management Group (DMG),
University of Toronto,
35 St George Street, M5S1A4, ON
Canada
Room: SF 3001F
Phone: 416-946-8027khandker.nurulhabib@utoronto.ca

## Abstract

The paper presents an empirical investigation of telecommuting frequency choices by post-secondary students in Toronto. It uses a dataset collected through a large-scale travel survey conducted on post-secondary students of four major universities in Toronto and it employs multiple alternative econometric modelling techniques for the empirical investigation. Results contribute on two fronts. Firstly, it presents empirical investigations of factors affecting telecommuting frequency choices of post-secondary students that are rare in literature. Secondly, it identifies better a performing econometric modelling technique for modelling telecommuting frequency choices. Empirical investigation clearly reveals that telecommuting for school related activities is prevalent among post-secondary students in Toronto. Around 80 percent of 0.18 million of the post-secondary students of the region, who make roughly 36,000 trips per day, also telecommute at least once a week. Considering that large numbers of students need to spend a long time travelling from home to campus with around 33 percent spending more than two hours a day on travelling, telecommuting has potential to enhance their quality of life. Empirical investigations reveal that car ownership and living farther from the campus have similar positive effects on the choice of higher frequency of telecommuting. Students who use a bicycle for regular travel are least likely to telecommute, compared to those using transit or a private car. Owning a transit pass has a stronger negative influence on the choice of high-frequency telecommuting than that of bicycle ownership. Overall, the paper provides clear evidence regarding the influence of different factors affecting the frequency of telecommuting by post-secondary students. This evidence can provide useful insights for developing strategies/policies that could influence post-secondary students' telecommuting choices.

## 1. Introduction

Telecommuting has been widely considered as an alternative to physical travel for work/school that has the potential to reduce commuting trips, especially during peak travel periods in urban areas. Therefore, telecommuting is considered as a travel demand management (TDM) policy that can help the urban in managing peak period traffic congestion, fossil fuel consumption, greenhouse gas emission and air pollution (Kim et al. 2015). Studies have shown that telecommuting might have an indirect influence on shaping urban forms and land use patterns (Ettema 2010). Mixed findings on this issue are evident. For example, Mokhtarian et al. (2004) found that telecommuters had lower total commute distances than non-commuters. Rhee (2007), in turn, found that telecommuting could encourage urban sprawl. However, from an individual's point of view, telecommuting is seen as enhancing the quality of life by reducing transportation costs, increasing time spent at home, and even improving work efficiency (Dissanayake 2008; Lari 2012; ScienceDaily 2013; Kim 2016).

Since the 1970s, a considerable amount of research has been dedicated to understanding the telecommuting behaviour of workers. In large urban areas, the most substantial impact of telecommuting should original from the telecommuting of full- or part-time workers. However, it is also true that a good portion of peak period travellers is post-secondary students making trips to school. In Canada, post-secondary students represent over 20 percent of labour force age category of 15 to 30-year-olds (Statistics Canada 2016). Also, with the rise of a knowledge-based economy, the participation of youths (employed or unemployed) in post-secondary education is increasing. Like any urban commuter, the travel distance (resulting travel time and cost) between homes to campus does matter to the post-secondary students. Previous studies showed that the home to campus travel distance even discourages participation in post-secondary education (Frenette 2002, 2003). Thus, telecommuting[1] by post-secondary students can have a significant influence on urban travel demands as well as students' quality of life.

In regional travel demand models, post-secondary students are not often separated from regular workers, and as a result, work/school trips are modelled together. However, post-secondary students are often under-represented in regional household travel surveys that are used for evidence-based regional transportation planning studies and policy developments. Thus the travel behaviour of post-secondary students is often overlooked or not properly understood. For example, in the Greater Toronto and Hamilton Area (GTHA) post-secondary students share over 17 percent of the total population. However, the latest regional household travel survey, from 2011, was a 5 percent sample household travel survey of the GTHA and had less than a 12 percent sample share of post-secondary students (TTS 2011). With an increasing rate of youth participation in post-secondary education programs (Marshall 2010), more empirical research on the travel behaviour of post-secondary students is deemed necessary. In this regard, investigating telecommuting choices of post-secondary students is of particular importance as telecommuting during post-secondary student life may also have an influence on their future telecommuting behaviour when they enter the workforce. Evidence showed that in Canada, more than 20 percent of workers in the workforce with post-secondary education telecommute (Turcott 2010). We are not aware of any study on telecommuting behaviour that focuses explicitly on post-secondary students.

[1] Herein, telecommuting by post-secondary students refers to studying or doing school related activities from home and it is maintained throughout this paper.

This paper contributes to this gap in the literature by presenting an empirical investigation of the factors that influence the telecommuting frequencies of post-secondary students in Toronto. The study makes use of a recently completed survey of post-secondary students of four major universities in Toronto, which is roughly an 8 percent sample survey representing more than 0.18 million post-secondary student population in the region. The survey was a travel diary survey that collected personal and household attributes of the students along with information on regular travel mode choices and weekly frequencies of telecommuting for school trips. The paper investigates alternative modelling techniques for investigating telecommuting frequency choices. The contribution of the article is of twofold. In essence, it presents an empirical investigation of telecommuting frequency choices by post-secondary students, and it evaluates alternative econometric models to identify an appropriate technique for modelling telecommuting frequency choices.

The remainder of the paper is organized as follows. Section 2 presents a brief overview of the literature on modelling telecommuting frequencies. Section 3 presents a discussion on the dataset used for empirical investigation. Section 4 presents alternative modelling techniques that are evaluated for empirical investigation. Section 5 presents a discussion on the empirical investigations. Finally, the paper concludes with a summary of key findings and recommendations for further research.

## 2. Literature Review

A number of studies in the literature investigated various aspects/impacts of telecommuting. Among the earliest studies, Mokhtarian (1978) speculated that the impact of telecommuting on reducing travel demand might be moderate to low due to the presence of many influential factors. Salomon (1986) classified three types of effects of telecommuting on travel demand: complementary; supplementary; and enhancing. Hamer et al. (1991) presented a before-after investigation of telecommuting effects and found that telecommuting could reduce overall travel demand as much as 17%. Mokhtarian et al. (1995) found telecommuting might not have any effect on non-work related travel demand. However, Mokhtarian and Salomon (1997) as well as Stanek and Mokhtarian (1998) investigated the preference for telecommuting and found that travel-related variables, such as commute distance and time, have insignificant effects on choosing to telecommute. Many such investigations are based on data analysis, often without developing any explicit econometric models.

Application of a linear or non-linear regression model for modelling travel demands (distance or vehicle-distance travel) is widespread in the literature to investigate the impact of telecommuting on travel demands. Choo et al. (2005), for example, used an aggregate time-series model and found that telecommuting had a supplementary relationship with total travel demand. However, Zhu (2012 and 2013), as well as Zhu and Mason (2014), used regression models to investigate multiple household travel survey data and found that telecommuting had complementary effects on daily travel demand. A recent study used a censored regression model for investigating effects of telecommuting on total household travel demand, and found that households with telecommuting head of households may even have increased travel demand (Kim 2016).

For a quantitative evaluation of the influence of various factors on the choice of telecommuting researchers have used econometric modelling techniques. Among many such empirical investigations a study by Bernardino and Ben-Akiva (1996) is the earliest that employed advanced econometrics to investigate jointly the probability of having the options of telecommuting along with the frequency of telecommuting if the option is available. They estimated a structural equation model for telecommuting options availability along with a binary choice model for telecommuting decision. They found telecommuting could increase productivity and improve the quality of life of workers. Various household and socio-economic variables are found to significantly influence telecommuting choices. Mannering and Mokhtarian (1995) estimated an unordered multinomial logit model of telecommuting frequency choice. They found that work statuses and work environment-related attributes play a critical role in defining the choices of telecommuting frequencies. Drucker and Khattak (2000) used ordered probit regression, a regression model with probit sample selection and unordered multinomial logit models to investigate the choice and frequencies of telecommuting. They found that ordered regression and unordered discrete choice models gave very similar results. In terms of influential factors, they found that the presence of children, gender, and the unavailability of free parking at the workplace play essential roles.

Bhat and Popuri (2003) used a binary choice model for modelling the choice of telecommuting, but the frequency of telecommuting is modelled as an ordered regression model. They estimated both models jointly to capture correlation between the choice of telecommuting and subsequent frequencies telecommuting. They found that individual and household demographics and work-related attributes were significant determinants of telecommuting adoption and frequency. Walls et al. (2007) used a similar approach, but they estimated two model components separately. Zhou et al. (2009) used an ordered regression model for investigating frequencies of telecommuting and found that as the distance between home and work increased, this would play a critical role in telecommuting frequencies, and thus a longer distance induced higher frequencies of telecommuting.

Similarly, Asgari et al. (2014) also used ordered regression for modelling telecommuting frequency choices and found telecommuting frequency is inherently related to a worker's lifestyle arrangements. As opposed to a simply ordered regression, Sing et al. (2013) used a generalized ordered regression model for investigating the option (availability), choice, and frequency of telecommuting. They highlighted the fact that a generalized ordered regression overcomes the limitations that a simply ordered regression does not allow for estimating separate covariate effects for different ordered categories. Although telecommuting frequencies are inherently ordered, a different level (frequency) of telecommuting may separate relationships with covariates. Moreover, they highlight the fact that the joint modelling of the option (supply side) in the telecommuting choice model is necessary to accurately capture the effects of different variables on telecommuting frequency choices.

In a most recent study, Paleti (2016) proposed a count variable choice model (as opposed to a count variable regression model) and used for modelling telecommuting frequency choices. The count variable choice model uses a closed-form mathematical structure, where a number of choices (frequencies) are explained through discrete count-variable distribution (Poisson or Negative binomial), but the alternative frequencies follow either a multinomial logit or an ordered probability model. This approach also allows separate count-specific covariate functions along with an underlying average rate function. This is a versatile econometric approach that combines properties of a count variable regression with a Random Utility Maximizing (RUM) discrete or ordered discrete choice model. The paper compared the alternative telecommuting frequency choice models and found that the proposed count variable choice model outperforms all other approaches. In this paper, Paleti's count variable choice model is compared against an ordered regression econometric model for modelling telecommuting frequency choices of post-secondary students in Toronto.

## 3. Description of Survey and Data for Empirical Investigation

Data for an empirical investigation came from a large-scale post-secondary student travel survey that was conducted in Toronto among 0.18 million post-secondary students of four major universities in Toronto (StudentMoveTO 2016). The survey was one of the first initiatives to promote evidence-based research on post-secondary students' travel behaviour in the region, which have largely been ignored. A total of four universities with seven campuses were targeted. These were: University of Toronto (downtown, Mississauga, and Scarborough campuses); OCAD (Ontario College of Art and Design) University (downtown Toronto); Ryerson University (downtown Toronto); and York University (Glendon and Keele campuses). The university locations cover Toronto and the York region, but the home locations of the post-secondary students of these campuses encompass the City of Toronto and beyond. In total, around 70 percent of students of these campuses live in the City of Toronto (30 percent of whom live in the downtown Toronto), and 30 percent live just outside the City of Toronto. The survey was conducted through a Web-based survey tool with an 8.3 percent completion rate (resulting in a sample of around 15,000 post-secondary students).

The preliminary data of the completed surveys reveal that these post-secondary students contribute to the region's travel demand by an average 36,000 trips per day and around 37 percent of these trips are of home to campus trips. The travel diary data collected preliminary data analysis, and it reveals that about 33 percent of all post-graduate students in Toronto spend more than two hours traveling to and from campus. Thus telecommuting for school activities has the potential to enhance students' quality of life. It is clear that post-secondary students in Toronto are well aware of this potential as around 80% of students reported telecommuting at least once a week. Figure 1 presents a number of ways of understanding the telecommuting behaviour of post-secondary students, along with some key summary statistics. It seems that the potential benefit of telecommuting for school activities is well understood by the student population under investigation. It also becomes clear that telecommuting is prevalent among all mode users, even those who walk or bicycle from home to campus. Telecommuting is also commonplace among graduate and undergraduate students.

| **Variables** | **Statistics by weekly frequencies of telecommuting** | | | | | | |
|---|---|---|---|---|---|---|---|
| | **0** | **1** | **2** | **3** | **4** | **5** | **6+** |
| Age (mean) | 23.14 | 23.00 | 22.83 | 23.26 | 24.85 | 27.20 | 26.27 |
| Number of female/Number of male | 1.71 | 1.73 | 2.06 | 2.48 | 2.96 | 2.44 | 1.56 |
| Home to campus distance: km | 12.67 | 11.48 | 11.58 | 13.55 | 15.03 | 13.71 | 12.51 |
| 18% of sample having a car | 19.9% | 15.9% | 25.4% | 17.7% | 10.5% | 5.6% | 5.0% |
| Age of acquiring driver's license (average years) | 18.68 | 18.69 | 18.67 | 19.91 | 19.03 | 19.63 | 18.91 |
| 47.62% of sample having a bicycle | 20.7% | 18.4% | 29.8% | 16.2% | 8.6% | 3.5% | 2.7% |
| 56.56% of sample having transit passes | 19.9% | 19.0% | 32.9% | 15.2% | 6.5% | 3.2% | 3.3% |
| 55.96% of sample living with parents/family | 19.9% | 17.0% | 30.1% | 18.3% | 8.5% | 3.4% | 2.6% |
| 10.62% of sample living alone | 22% | 16.9% | 29.0% | 13.6% | 8.1% | 4.8% | 5.5% |
| 19.10% of sample living with someone of same generation | 19.9% | 19.7% | 29.6% | 14.3% | 7.8% | 4.8% | 3.8% |
| Home to nearest bus stop (average, km) | 0.27 | 0.26 | 0.26 | 0.28 | 0.29 | 0.31 | 0.31 |
| Home to nearest rail station (average, km) | 3.6 | 3.6 | 3.6 | 3.8 | 3.9 | 3.7 | 3.8 |
| Home to nearest subway station (average, km) | 7.9 | 7.2 | 7.3 | 9.3 | 9.31 | 7.5 | 8.0 |
| Area (sq. km) of 1 km walking buffer | 1.41 | 1.43 | 1.42 | 1.36 | 1.36 | 1.41 | 1.44 |

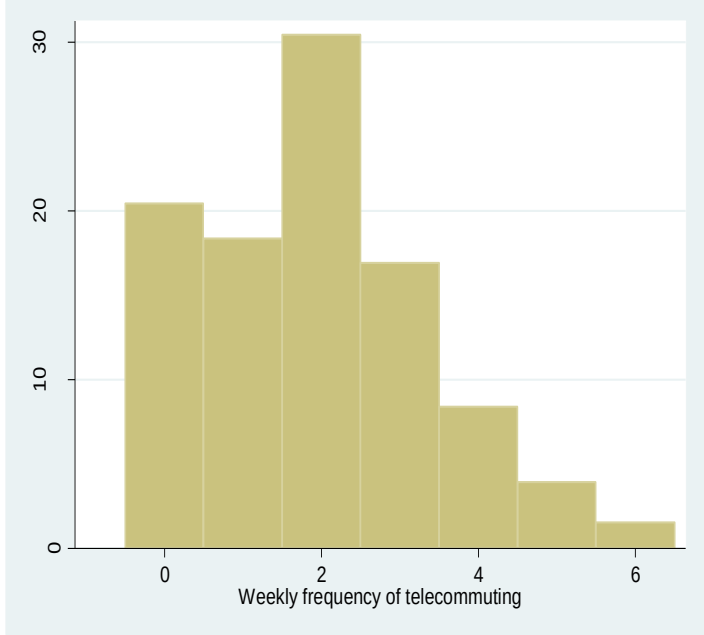

Overall sample distribution telecommuting frequencies

Cleaned sample size: 13528

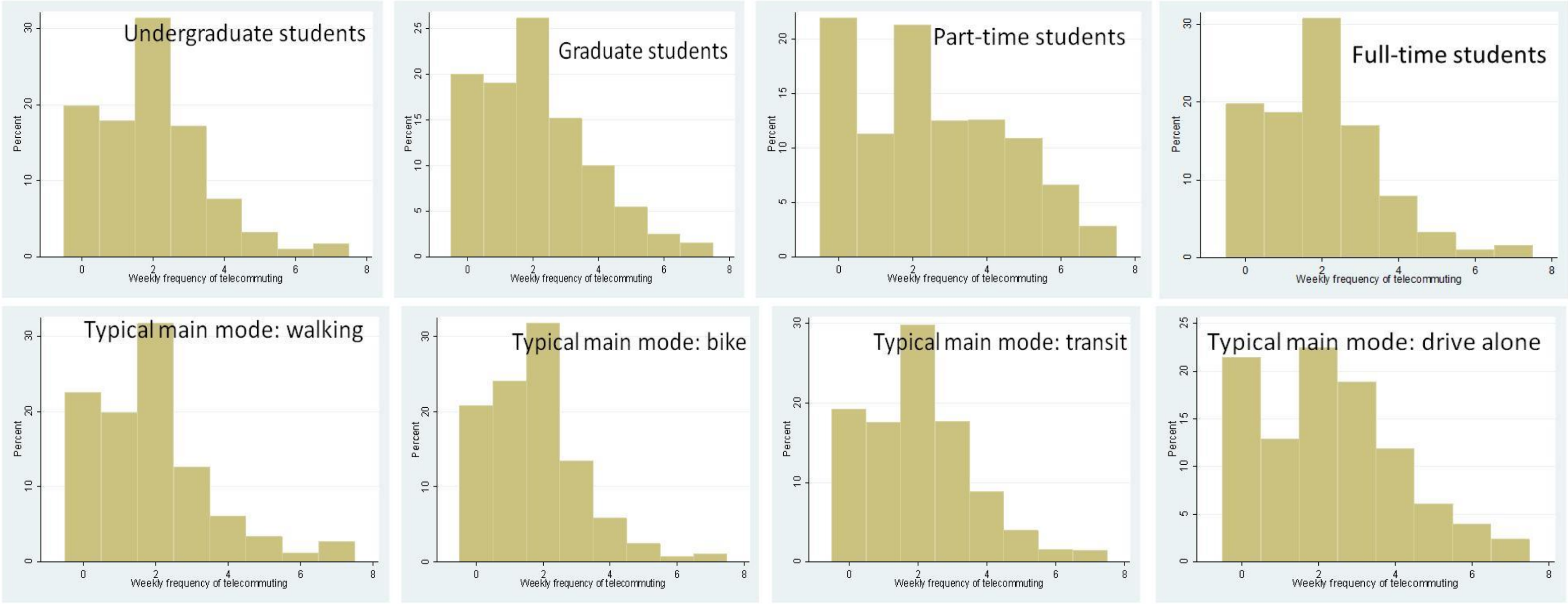

**Figure 1: Summary statistics.**

As shown in Figure 1, there are distributions of weekly frequencies of telecommuting by post-secondary students in Toronto. This study found that it is difficult to ascertain the role of different variables on the rate of telecommuting as the relative proportion and distribution vary widely with respect to different variables. It is also clear that, in general, around 20 percent of students do not telecommute. Consequently, data on telecommuting frequencies are somewhat 'zero-inflated' and empirical investigations need to accommodate this fact. In addition, the distribution of weekly telecommuting frequencies shown in Figure 1 also indicates that there is a wide range of 'heterogeneity' in telecommuting frequency choices. As a result, the empirical investigations need to accommodate these issues.

## 4. Econometric Models

This paper uses data on telecommuting (weekly frequency) collected in the survey for empirical investigation. Considering the intention of accommodating zero-inflation and preference heterogeneity of weekly telecommuting frequencies and based on understandings of different modelling techniques used by various previous studies, two alternative econometric approaches are used in this investigation. The first approach is ordered regression. However, as opposed the normal regression approach, an Ordered Extreme Value (OEV) approach is used as this allows for the accommodation of heterogeneity (over-dispersion of telecommuting frequency data) through a Gamma parametric mixing distribution. A further specification of the OEV is developed that allows for a separate model for zero frequency telecommuting that addresses the zero-inflation of telecommuting frequency data. For the second approach, a RUM-based count variable choice model is used that follows the proposed formulations of Paleti (2016). It models the underlying mean frequency of telecommuting through a Negative Binomial (NB) or Poisson process and the choices among distinct alternative frequencies are modelled as the Ordered Generalized Extreme Value (OGEV) process. Hence, two alternative formulations are named as the Mean frequency NB-OGEV model and Mean frequency Poisson-OGEV model. In total, four different types of econometric models are estimated, and the results are compared. The econometric formulations of the models are explained below.

### 4.1: OEV model and split population OEV model

The OEV approach assumed an underlying continuous function of telecommuting frequency as:

$$F_i = U_i + \varepsilon_i \tag{1}$$

Here, the observed frequency id $f_i$

Underlying continuous function of the counterpart if $f_i$ is $F_i$

$\varepsilon_i$ is a random error component

The probability, y, of an observed frequency, $f_i = k$:

$$\Pr(f_i = k) = G\left(\delta_k - \sum \beta x_i\right) - G\left(\delta_{k-1} - \sum \beta x_i\right) \tag{2}$$

Here, $\delta_{k-1}$ and $\delta_k$ are the nonparametric baseline frequency threshold for (k-1) and k

G(.) is the cumulative distribution function of the distribution of $\varepsilon_i$

With the assumption of an extreme value distribution for the random term $\varepsilon_i$, the probability of observed frequency becomes:

$$\Pr(f_i = k) = \left(1 - e^{\left(-e^{(\delta_k - \Sigma\beta x_i)}\right)}\right) - \left(1 - e^{\left(-e^{(\delta_{k-1} - \Sigma\beta x_i)}\right)}\right) \quad (3)$$

This assumption does not consider any dispersion or pattern of heterogeneity. A common practice of inducing a parametric dispersion/heterogeneity is to mix a positive distribution (Han and Hausman 1990; Bhat 1996). Considering mixing a Gamma distribution of unit mean and a $\sigma^2$ variance, the probability equation becomes:

$$\Pr(f_i > k) = \left(1 + \frac{1}{\sigma^2}\Delta_k e^{(-\Sigma\beta x_i)}\right)^{-\sigma^2} \quad (4)$$

$$\Pr(f_i < k) = 1 - \left(1 + \frac{1}{\sigma^2}\Delta_k e^{(-\Sigma\beta x_i)}\right)^{-\sigma^2} \quad (5)$$

$$\Pr(f_i = k) = \left(1 + \frac{1}{\sigma^2}\Delta_{k-1} e^{(-\Sigma\beta x_i)}\right)^{-\sigma^2} - \left(1 + \frac{1}{\sigma^2}\Delta_k e^{(-\Sigma\beta x_i)}\right)^{-\sigma^2} \quad (6)$$

For the maximum *K* frequencies, a total of (K-1) non-parametric base line thresholds can be identified. So,

$$\Pr(f_i = 0) = 1 - \left(1 + \frac{1}{\sigma^2}\Delta_k e^{(-\Sigma\beta x_i)}\right)^{-\sigma^2} \quad (7)$$

$$\Pr(f_i = K) = \left(1 + \frac{1}{\sigma^2}\Delta_k e^{(-\Sigma\beta x_i)}\right)^{-\sigma^2} \quad (8)$$

$$\Pr(f_i = k) = \left(1 + \frac{1}{\sigma^2}\Delta_{k-1} e^{(-\Sigma\beta x_i)}\right)^{-\sigma^2} - \left(1 + \frac{1}{\sigma^2}\Delta_{k-1} e^{(-\Sigma\beta x_i)}\right)^{-\sigma^2} \quad (9)$$

Considering all possible non-negative values, the likelihood functions of any observation:

$$L_i = \begin{cases} \left[1 - \left(1 + \frac{1}{\sigma^2}\Delta_1 e^{(-\Sigma\beta x_i)}\right)^{-\sigma^2}\right] & \text{if } f_i = 0, \text{ no telecommuting} \\ \left[\left(1 + \frac{1}{\sigma^2}\Delta_{k-1} e^{(-\Sigma\beta x_i)}\right)^{-\sigma^2}\right] - \left[\left(1 + \frac{1}{\sigma^2}\Delta_k e^{(-\Sigma\beta x_i)}\right)^{-\sigma^2}\right] & \text{if } f_i = k, \text{ any intermediate frequency} \\ \left[\left(1 + \frac{1}{\sigma^2}\Delta_K e^{(-\Sigma\beta x_i)}\right)^{-\sigma^2}\right] & \text{if } f_i = K, \text{ the maximum frequency} \end{cases} \quad (10)$$

For this formulation, marginal effect of any variable (x) becomes:

$$ME_x = \begin{cases} -\left[\left(1+\frac{1}{\sigma^2}\Delta_1 e^{(-\Sigma\beta x_i)}\right)^{-\sigma^2-1}\right]\beta\Delta_1 e^{(-\Sigma\beta x_i)} & if \quad f_i = 0 \\ \left[\left(1+\frac{1}{\sigma^2}\Delta_{k-1} e^{(-\Sigma\beta x_i)}\right)^{-\sigma^2-1}\right]\beta\Delta_{k-1} e^{(-\Sigma\beta x_i)} & \\ \quad -\left[\left(1+\frac{1}{\sigma^2}\Delta_k e^{(-\Sigma\beta x_i)}\right)^{-\sigma^2-1}\right]\beta\Delta_k e^{(-\Sigma\beta x_i)} & if \quad f_i = k\text{, any intermediate value} \\ \left[\left(1+\frac{1}{\sigma^2}\Delta_K e^{(-\Sigma\beta x_i)}\right)^{-\sigma^2-1}\right]\beta\Delta_K e^{(\delta_K-\Sigma\beta x_i)} & if \quad t_i = K\text{, the maximum frequency} \end{cases} \tag{11}$$

The predicted baseline threshold at any interval $k$ is based on estimated non-parametric baseline δs are estimated:

$$\lambda_0(k)' = \Pr\left(D_i \in [d_{k-1}, d_k] \mid D_i \geq d_{k-1}\right) = \left(\Delta_k - \Delta_{k-1}\right) \tag{12}$$

Equation (9) does not treat zero frequency separately from any other positive frequency values. However, considering that the dataset may have relatively higher numbers of observations with zero frequency than other positive frequencies, a special treatment of zero frequency choice needs to be considered:

$$L_i = \left(\left(1-F\left(\sum\gamma z_i\right)\right)L_i\right)^{1-C} \times \left(F\left(\sum\gamma z_i\right)\left[1-\left(1+\frac{1}{\sigma^2}\Delta_K e^{(-\Sigma\beta x_i)}\right)^{-\sigma^2}\right]\right)^C \tag{13}$$

here,
$C$ is an indicator that takes the value of 1 if the individual does telecommute and 0 otherwise
F( . ) is logit probability of not telecommuting at all.
$L_i$ is the probability of non-zero frequencies as explained in equation (10)

This is a latent frequency model with zero-inflation. For this formulation the marginal effect of any variable ($z$) on probability of not telecommuting is:

$$ME_z(\text{Not acquiring}) = \left(F\left(\sum\gamma z_i\right)\left(1-F\left(\sum\gamma z_i\right)\right)\right)\gamma \tag{14}$$

Marginal effect of any variable (x) on non-zero frequency:

$$
ME_x = \begin{cases} -\left(1-F(\gamma z)\right)\left[\left(1+\frac{1}{\sigma^2}\Delta_1 e^{(-\Sigma\beta x_i)}\right)^{-\sigma^2-1}\right]\beta\Delta_1 e^{(-\Sigma\beta x_i)} & if \quad f_i = 1 \\ \left(1-F(\gamma z)\right)\left[\left(1+\frac{1}{\sigma^2}\Delta_{k-1} e^{(-\Sigma\beta x_i)}\right)^{-\sigma^2-1}\right]\beta\Delta_{k-1} e^{(-\Sigma\beta x_i)} & \\ -\left(1-F(\gamma z)\right)\left[\left(1+\frac{1}{\sigma^2}\Delta_k e^{(-\Sigma\beta x_i)}\right)^{-\sigma^2-1}\right]\beta\Delta_k e^{(-\Sigma\beta x_i)} & if \quad f_i = k\text{, any intermediate value} \\ \left(1-F(\gamma z)\right)\left[\left(1+\frac{1}{\sigma^2}\Delta_K e^{(-\Sigma\beta x_i)}\right)^{-\sigma^2-1}\right]\beta\Delta_K e^{(\delta_K-\Sigma\beta x_i)} & if \quad t_i = K\text{, the maximum frequency} \end{cases} \tag{15}
$$

## 4.2: Mean frequency NB-OGEV and mean frequency Poisson-OGEV model

Following Paleti (2016), for the RUM-based Ordered GEV model with Negative Binomial (NB) count choices, the probability of any observed choice of telecommuting frequency, *y*:

$$
\Pr(y) = \left(\frac{r}{r+\lambda}\right)^r \frac{\Gamma(r+y)}{\Gamma(y+1)\Gamma r}\left(\frac{\lambda}{r+\lambda}\right)^y \tag{16}
$$

here,
*y* is the number of trips/usage
*r* is a non-negative dispersion parameter
*λ* is the expected number of trips/usage = exp(∑βx)
Further elaboration:

$$
\Pr(y) = \frac{\left(\frac{r}{r+\lambda}\right)^r \frac{\Gamma(r+y)}{\Gamma(y+1)\Gamma r}\left(\frac{\lambda}{r+\lambda}\right)^y}{\sum_{y=0}^{Y}\left(\left(\frac{r}{r+\lambda}\right)^r \frac{\Gamma(r+y)}{\Gamma(y+1)\Gamma r}\left(\frac{\lambda}{r+\lambda}\right)^y\right)} = \frac{\frac{\Gamma(r+y)}{\Gamma(y+1)\Gamma r}\left(\frac{\lambda}{r+\lambda}\right)^y}{\sum_{y=0}^{Y}\left(\frac{\Gamma(r+y)}{\Gamma(y+1)\Gamma r}\left(\frac{\lambda}{r+\lambda}\right)^y\right)} \tag{17}
$$

$$
\Pr(y) = \frac{\exp(V_y)}{\sum_{y=0}^{Y}\exp(V_y)}, \quad \text{Here } V_y = \ln\left(\frac{\Gamma(r+y)}{\Gamma(y+1)\Gamma r}\left(\frac{\lambda}{r+\lambda}\right)^y\right) \tag{18}
$$

This is a multinomial logit (MNL) version of Negative Binomial Choice model. We can further add count-specific elements to the systematic utility to capture over/underdispersion of any specific counts:

$$
\Pr(y) = \frac{\exp(V_y + \eta_y)}{\sum_{y=0}^{Y}\exp(V_y + \eta_y)} \tag{19}
$$

here, $\eta$ is an additional systematic count-specific systematic heterogeneity over and above NB mean frequency and is can be further explained as a linear-in-parameter function of variable set b and corresponding coefficients, $\omega$; $\sum \omega b$ .

An additional count-specific component adds count-specific systematic heterogeneity over and above NB distribution of frequency of mean frequency. Therefore, this model is referred as a Mean frequency NB model. As identified by Paleti (2016), the NB distributing collapses into a Poisson distribution for a large value of *r*. In the case of additional count-specific constants, often *r* becomes too large to retain the NB formulation. So with a Poisson formulation, the probability function becomes:

$$\Pr(y) = \frac{e^{-\lambda}\lambda^{y}}{y!} \tag{20}$$

Here,
*y* is the number of trips/usage
*r* is a non-negative dispersion parameter
$\lambda$ is the expected number of trips/usage = exp(∑βx)

By using Taylor's series approximation:

$$\Pr(y) = \frac{\lambda^{y}/y!}{\exp(-\lambda)} = \frac{\lambda^{y}/y!}{\sum_{y=0}^{Y}(\lambda^{y}/y!)} = \frac{\exp(\ln(\lambda^{y}/y!))}{\sum_{y=0}^{Y}\exp(\ln(\lambda^{y}/y!))}, \quad \text{Here } V_{y} = \ln\left(\lambda^{y}/y!\right) \tag{21}$$

This is an MNL version of a Poisson Choice Model. Similar to equation (18), we can further add count-specific elements to the systematic utility to capture over/under dispersion of any specific counts:

$$\Pr(y) = \frac{\exp(V_{y} + \eta_{y})}{\sum_{y=0}^{Y}\exp(V_{y} + \eta_{y})} \tag{22}$$

here, $\eta$ is an additional systematic count-specific systematic heterogeneity over and above Poisson mean frequency and it can be further explained as a linear-in-parameter function of variable set b and corresponding coefficients, $\omega$; $\sum \omega b$ .

Both Mean frequency NB and Poisson formulations explained above considered an MNL formulation that assumes that the choice of consecutive frequencies (*y-1*, *y*, *y+1*) are independent of each other. Considering a serially ordered correlation between successive frequency choices, we can assume an Ordered Generalized Extreme Value (OGEV) model assumption as following (Small 1987; Paleti 2016):

$$\Pr(y) = \frac{e^{\left(\frac{(V_y+\eta_y)}{\rho}\right)}\left[\left(e^{\left(\frac{(V_{y-1}+\eta_{y-1})}{\rho}\right)} + e^{\left(\frac{(V_y+\eta_y)}{\rho}\right)}\right)^{\rho-1} + \left(e^{\left(\frac{(V_y+\eta_y)}{\rho}\right)} + e^{\left(\frac{(V_{y+1}+\eta_{y+1})}{\rho}\right)}\right)^{\rho-1}\right]}{\sum_{r=0}^{R+1}\left(e^{\left(\frac{(V_r+\eta_r)}{\rho}\right)} + e^{\left(\frac{(V_{r-1}+\eta_{r-1})}{\rho}\right)}\right)^{\rho}} \tag{23}$$

$$where\ \ e^{\left(\frac{(V_r+\eta_r)}{\rho}\right)} = 0 \ \ for\ r<0\ and\ r>R$$

Marginal effect of the OGEN model:

$$= \sum_{O} \Pr(O)\Pr(k \mid O)\left[(1-\Pr(k)) + \left(\frac{1}{\rho}-1\right)(1-\Pr(k \mid O))\right]\left(\frac{\partial V_k}{\partial x}\right)$$

$$= \left[\begin{array}{l} \Pr(k-1,k)\Pr(k \mid k-1,k)\left[(1-\Pr(k)) + \left(\frac{1}{\rho}-1\right)(1-\Pr(k \mid k-1,k))\right] \\ + \Pr(k,k+1)\Pr(k \mid k,k+1)\left[(1-\Pr(k)) + \left(\frac{1}{\rho}-1\right)(1-\Pr(k \mid k,k+1))\right] \end{array}\right]\left(\frac{\partial V_k}{\partial x}\right) \tag{24}$$

Where, O indicates the clusters of pairs

For the NB formulation, it becomes the Mean frequency NB-OGEV:

$$V_k = V_y = \ln\left(\frac{\Gamma(r+y)}{\Gamma(y+1)\Gamma r}\left(\frac{\lambda}{r+\lambda}\right)^y\right) = \ln\left(\frac{\Gamma(r+y)}{\Gamma(y+1)\Gamma r}\left(\frac{\exp(\beta x)}{r+\exp(\beta x)}\right)^y\right)$$

$$\begin{aligned} So,\ \frac{\partial V_y}{\partial x} &= \left(\frac{\Gamma(r+y)}{\Gamma(y+1)\Gamma r}\left(\frac{\exp(\beta x)}{r+\exp(\beta x)}\right)^y\right)^{-1} \times \frac{\Gamma(r+y)}{\Gamma(y+1)\Gamma r} \times y\left(\frac{\exp(\beta x)}{r+\exp(\beta x)}\right)^{y-1} \\ &\quad \times \frac{(r+\exp(\beta x))\exp(\beta x)\beta - \exp(\beta x)\exp(\beta x)\beta}{(r+\exp(\beta x))^2} \\ &= y\beta\left(1-\frac{\lambda}{r+\lambda}\right) = y\beta\left(\frac{r}{r+\lambda}\right) \end{aligned} \tag{25}$$

For the Poisson formulation it becomes the Mean frequency Poisson OGEV formulation:

$$\begin{aligned} &V_y + \eta_y = \ln(\lambda^y / y!) + \eta_y = y\ln(e^{\sum \beta x}) - \ln(y!) + \eta_y = y(\sum \beta x) - \ln(y!) + \eta_y \\ &So,\ \frac{\partial(V_y+\eta_y)}{\partial x} = y\beta,\ where\ \lambda = e^{\sum \beta x} \end{aligned} \tag{26}$$

In case of both NB and Poisson formulation, the marginal effect of covariates in additional utility component:

$$\frac{\partial(V_y+\eta_y)}{\partial b}=\omega, where\ \eta_y=\sum\omega b \tag{27}$$

All of these models have closed form formulations and can be estimated by using classical estimation techniques. In this investigation, programs are developed in GAUSS using its MAXLIK module to estimate through the maximum likelihood estimation technique (Aptech 2016).

## 5. Empirical Investigation

A total of four models are estimated, and these are: (1) Mean frequency NB-OGEV model; (2) Mean Frequency Poisson OGEV model; (3) Ordered Extreme Value (OEV) with Gamma heterogeneity model; and (4) Split population OEV with Gamma heterogeneity model. Akaike Information Criteria (AIC) and Bayesian Information Criteria (BIC) are used for comparing the relative performances of the four types of statistical models:

AIC = 2 x (Number of estimated parameters) – 2 x log-likelihood at convergence
BIC = 2 x log-likelihood at convergence
– (Number of estimated parameters x log of the number of observations)

Both AIC and BIC measures the performance of a statistical model in terms of fitting the underlying behavioural process of actually observed data generation processes. In both cases, the lower the value means, the better the statistical model is in mimicking the actual underlying behavioural processes that it is intended to model. Following AIC and BIC measures, the goodness-of-fit of the models are compared by estimating *Rho-square* value:

$$Rho-Squared=1-\frac{\text{Log}\,likelihood\,\text{at convergence}}{\text{Log}\,likelihood\,\text{of a null model}}$$

### 5.1 Performances of different econometric approaches in modelling telecommuting frequencies of post-secondary students

A summary of the estimated model parameters is presented in Table 1. It is interesting to note that not all models can accommodate the same set of covariates. The split population OEV model with Gamma heterogeneity gives the best statistical performance in terms of lowest AIC and BIC values as well as highest *Rho-squared* value. However, the main reason for retaining all four models is that their statistical performances are not drastically different from each other (the AIC, BIC, and *Rho-squared* values are close), and not all models can explain the roles of all variables that may influence the choice of telecommuting. The close performances of the different econometric formulations containing different sets of covariates indicate the complicated behavioural process involved in telecommuting choice.

Between the RUM-based count variable models, the NB model slightly outperforms (through better statistical fit and a higher number of estimates with 95 percent confidence) the Poisson model indicating that there is substantial heterogeneity in the telecommuting frequency choices of post-secondary students. This is proven by the value of the dispersion parameter, which is greater than 1.

Among the OEV models with Gamma heterogeneity, the split population version outperforms the other. Modelling choice of no telecommuting is synonymous with a zero-inflation model. In the empirical investigation of Paleti (2016), it seems that the RUM-based count variable model outperformed the zero-inflated count regression model. The RUM-based count variable choice model can have a systematic count-specific utility component that works similar to a zero inflation component. However, in this investigation, it is found that accommodating zero-inflation in the form of a split population with the OEV model can outperform the RUM-based approach. Paleti (2016) presented a RUM-based count variable with an additional county-specific systematic component only for the multinomial logit model. However, in this investigation, additional county-specific systematic components were accommodated within an OGEV formulation. The split population OEV model seems to indicate that the choice of not telecommuting has a very different underlying behavioural process than the choice of the non-zero frequency of telecommuting. Therefore, even though the RUB-based model can have an extra count-specific utility component, the explicitly zero-frequency specific function of the split population OEV gives better performance.

In the following discussion, the two final competing models: the RUB-based NB count variable choice model and the split population OEV model are discussed together. To further highlight the relative influences of different variables in each model, the estimated marginal effects of all variables of the best-performing model, the split-population OEV model is presented in Figure 2. The marginal effects of the covariates of the Mean frequency NB-OGEV model are also estimated and discussed along with those of split population OEV model but are not presented in figure or table format to save space.

The RUM-based count variable choice model could have only two variables for the systematic explanation of the choice of not telecommuting among post-secondary students in Toronto. The most influential one among these is the dummy indicating 'living on-campus'. This choice is reasonable, given that students who live on-campus are least likely to telecommute. However, the split population OEV model reveals that students living on campus are most likely not to telecommute and at least likely to telecommute 1 or 2 times a week, but may also telecommute more than two times a week. As opposed to only two variables accommodated by the RUM-based approach, the split population OEV model accommodates the effects of: age; gender; student status; bicycle ownership; transit pass ownership; regular travel modes; changes in mode choice between winter and other times of the year; housing type; living situation and the land use attributes of home location on the choice of telecommuting. These effects are discussed in the context of different frequencies of weekly telecommuting in the following section.

### 5.2 Role of personal and household attributes on telecommuting frequencies

Age of the individual seems to be the most influential variable in explaining the choices of telecommuting frequencies that are consistently captured by both modelling approaches. The RUM-based model reveals that with increasing age the probability of telecommuting increases from 1 to 2 times a week and then decreases for higher frequencies. However, the split population OEV model reveals a more complicated effect of age. It shows that older students are least likely to telecommute once a week, but are more likely to telecommute more than thrice a week. Gender seems to play a different role in telecommuting choices for post-secondary students. The RUM-based model reveals that females are more likely to telecommute than males, but the gender effect shows a lognormal pattern of marginal effects with increasing frequencies of telecommuting. This is the case with RUM-based models as the split population OEV model indicates that female students are less likely to telecommute and will do so less than twice per week. However, female students are more likely to make more than twice per week than males. Interesting to note that graduate students are less likely to telecommute than undergraduate students (after controlling for age), but this effect is only captured by the RUM-based model. The RUM-based model reveals that students who received their driver's license later than the other students are less likely to telecommute than others. However, the OEV model reveals that students who got their driver's license later than others are more likely to telecommute once or twice per week, but less likely to telecommute more than twice a week than others who got their license earlier.

In terms of the type of housing, the students live in this study found that living in a condominium or an apartment off-campus shows a significant effect on telecommuting choices, as opposed to no effect for other housing types, e.g., detached or semi-detached house. Both models show that apartments/condominium dwellers are more likely to telecommute than those living on-campus or in other types of homes. Such types of dwelling are usually situated in urban cores or places closer to transit services. Perhaps better transit accessibility encourages students to participate in different activity types and thereby increases the likelihood of telecommuting.

### 5.3 Role of mobility tool ownership, regular mode choices and living arrangements

Mobility tool ownership, in terms of ownership of the car, transit pass, bicycle, car sharing membership, and so forth, affects the telecommuting choices of post-secondary students. The RUM-based model indicates that students who own cars, for example, are more likely not to telecommute at all and least likely to telecommute once or twice a week. However, it also shows that students living farther from campus are more likely to have higher frequencies of telecommuting. On the other hand, the split population OEV model consistently reveals that students who own cars have similar telecommuting behaviour as the students who live far from campus. It shows that car ownership and living far from campus influence higher frequencies of telecommuting (more significant than once or twice a week). This reveals that students living close to the campus but having access to a car would have similar patterns of telecommuting frequencies as those students living far from the campus and who do not have access to a car. Similarly, it can be said that students living far from the campus and have access to a car would have very high frequencies of telecommuting.

The effect of car sharing membership is not captured by the RUM-based model, but the OEV model reveals that students who own a car-sharing membership are least likely to telecommute once a week and are more likely to telecommute more than twice a week. Similar to car ownership, the RUM-based model explains that transit pass ownership decreases the odds of not telecommuting at all, while the OEV model reveals that carsharing membership negatively influences the odds of not telecommuting, as well as doing so once or twice a week (as opposed to telecommuting more often). Similar patterns are observed for bicycle ownership. However, both models reveal that the relative impact of transit pass ownership on telecommuting choices is higher than that of either bicycle ownership or car sharing.

Regular travel mode choice is shown to have profound effects on the choice of weekly telecommuting frequencies. Students who typically use a bicycle are least likely to telecommute, relative to those who use transit or a private car, for instance. The RUM-based model also reveals that students who use the same travel mode in winter and non-winter seasons are least likely to telecommute.

In terms of living arrangements, the RUM-based model reveals that post-secondary students who do not live with parents are more likely to telecommute than those living with parents. This makes sense, given these students are also older and more likely to be. However, the OEV model better captures the differential effects of different living arrangements. Post-secondary students living with a partner are the least likely to telecommute, while students living alone or living with partners are equally unlikely to telecommute once or twice a week.

### 5.4 Role of home location and land use attributes of home locations

Places of residence for post-secondary students have a strong influence on telecommuting behaviour in Toronto. The RUM-based model reveals that students living in downtown Toronto are less likely to telecommute than those living outside downtown Toronto. The OEV model indicates that living in the downtown area encourages telecommuting a maximum of once or twice a week, but discourages higher frequencies of telecommuting. Home zone population density does not show any effect on telecommuting behaviour of post-secondary students in Toronto. However, the OEV model captures that higher employment density at the home location increases the probability of not telecommuting at all. Home zone transit accessibility, measured in terms of distance from home to the nearest bus stop, rail station (GO rail) and subway station show strong influences on telecommuting behaviour for post-secondary students in Toronto: poor bus network accessibility increases the probability of telecommuting, while the opposite effect is true for rail and subway network accessibility. This means students living farther away from a subway or rail network are less likely to telecommute.

## 6. Conclusions and recommendations for further research

The contribution of this paper is twofold. First, it investigates the telecommuting of post-secondary students, which is rarely studied and poorly understood. Secondly, it employed multiple econometric modelling approaches to identify the approach that explains telecommuting behaviour the best. The paper also demonstrates the performance of a recently developed modelling approach for telecommuting choice (random utility maximizing count variable choice model) with respect to an alternative modelling approach of split population ordered extreme value model with Gamma heterogeneity. It reveals that in the case of post-secondary students in Toronto, a split population ordered extreme value modelling approach gives a better fit to observed data and a better explanation of the roles of covariates in explaining telecommuting choices. In terms of covariate effects, marginal effects of the same variable are not directly comparable as the probability formulations are different. In some cases, the OEV model provides greater distinguishing effects of some variables on different frequencies of telecommuting than the RUM-based model does. The effects of age and gender are two examples. Also, in some cases, both models complement each other in terms of the impact of some variables. The effect of the number of years in the university is another example. The OEV model captures that greater number of years student studies in university decreases the probability of not telecommuting and the RUM-based model explains that the greater number of years in university increases the probability of higher frequencies of telecommuting. Thus, for understanding the effects of different influential factors on the choices of telecommuting by post-secondary students, the paper highlights the importance of applying multiple modelling techniques.

Overall, both models reveal that older and female post-secondary students are more likely to telecommute than others. Car ownership and living off-campus also increase the probability of telecommuting. Students who use a bicycle for regular travel are the least likely to telecommute. Modal behaviour, in terms of not changing a travel mode in winter and non-winter seasons, also has an influence on telecommuting choice. Stable modal behaviour (same modes for all season) reduces the probability of telecommuting. In terms of mobility tool ownership, transit pass ownership has the highest negative impact on the choice of high frequencies of telecommuting frequencies compared to vehicle ownership, bicycle access or car-sharing membership. The city has a discounted transit pass policy for post-secondary students. Thus it seems that this policy may have an influence on discouraging post-secondary students from telecommuting and encouraging being on campus more regularly.

General household attributes, for example, household size, household car ownership, household income, and so forth, do not seem to have any influence on the telecommuting behaviour of post-secondary students in Toronto. Living arrangements, however, have a strong impact on their telecommuting choices.

Compared to living with parents or immediate family, living with partners, living alone and living with roommates seems to have very different effects on telecommuting frequencies. Land use attributes and transit accessibility of home locations play a critical role in influencing the telecommuting choices of post-secondary students in Toronto. In terms of transit network accessibility, bus, as well as rail and subway network accessibility, seems to have the opposite effect on the telecommuting behaviour of post-secondary students in Toronto. It is interesting to note that better bus and streetcar network accessibility increases telecommuting probability. However, better rail and subway network accessibility decreases the likelihood of telecommuting.

The cost/benefit of telecommuting by post-secondary students is complex. As per the literature on the benefit of telecommuting, it is clear that the choice of telecommuting would help students save time from travelling that can be used for other activities along with saving the students from the externalities of the transportation system, e.g., exposure to air pollution, traffic accidents, etc. At the same time, a high frequency of telecommuting may also deprive the students of campus life/environment. The result of the empirical investigation of this paper can be used for evaluating various policy relevancies. In general, the results of this investigation provide a clear understanding of the telecommuting behaviour of the post-secondary student population in Toronto that may be useful for university administrations in understanding/facilitating telecommuting of the students. From a general transportation planning and policy analysis perspective, these findings also have policy relevancies. For example, the City of Toronto has a discounted transit pass policy for post-secondary students. It seems that this policy has a negative influence on higher weekly frequencies of telecommuting on the 0.18 million post-secondary students in Toronto. There have been continuous efforts in strategy development by the City of Toronto to encourage bicycling and increasing bicycle ownership in the city. It seems that such strategies will also have a negative influence on higher weekly frequencies of telecommuting by post-secondary students in the city.

**Acknowledgement**
The research was funded by an NSERC Discovery Grant. The authors acknowledge the work of the StudentMoveTO survey team, especially Christopher Harding, for facilitating access to the disaggregated data for this study. The authors also acknowledge the help of Dr. Steven Farber for sharing parcel-level land use and transit accessibility data of the study area. However, all comments and interpretations presented in the paper are of the authors alone.

## Table 1: Empirical models

| | Mean frequency Poisson OGEV | Mean frequency NB OGEV | Covariate function of ordered EV | Split Population ordered EV |
|---|---|---|---|---|
| Total number of observations | 13528 | 13528 | 13528 | 13528 |
| Loglikelihood at convergence | -23728 | -23573 | -23523 | -23418 |
| Loglikelihood of null model | -28131 | -28131 | -28131 | -28131 |
| AIC Value | 47518 | 47209 | 47105 | 46924 |
| BIC Value | 47751 | 47442 | 47323 | 47254 |
| *Rho-squared* value | 0.157 | 0.162 | 0.164 | 0.168 |

| | **Mean frequency Poisson OGEV** | | **Mean frequency NB OGEV** | | **Covariate function of ordered EV** | | **Split Population ordered EV: Binary logit** | | **Split Population ordered EV: Covariate function** | |
|---|---|---|---|---|---|---|---|---|---|---|
| **Variable** | **Parameter** | ***t*-stat** | **Parameter** | ***t*-stat** | **Parameter** | ***t*-stat** | **Parameter** | ***t*-stat** | **Parameter** | **t-stat** |
| Constant | -0.241 | -1.59 | -0.811 | -5.01 | x | | -1.347 | -2.63 | | |
| Female | 0.089 | 7.32 | 0.117 | 7.19 | 0.196 | 7.11 | -0.225 | -4.99 | 0.195 | 5.82 |
| Log of Age | 0.337 | 10.61 | 0.448 | 9.62 | 0.782 | 10.35 | -0.314 | -2.62 | 0.757 | 8.68 |
| Graduate student | -0.027 | -1.72 | -0.049 | -2.25 | -0.059 | -1.72 | 0.075 | 1.28 | x | |
| Log of years in university | 0.014 | 1.57 | 0.012 | 1.05 | 0.038 | 2.00 | -0.073 | -2.18 | x | |
| Age of acquiring driver’s license | -0.009 | -1.91 | -0.017 | -2.82 | -0.020 | -2.16 | x | | -0.029 | -2.36 |
| Owning a car | x | | x | | -0.058 | -1.11 | x | | 0.054 | 1.16 |
| Member of a car sharing service | x | | x | | x | | x | | -0.091 | -1.45 |
| Transit pass owner | -0.107 | -8.00 | -0.118 | -6.71 | -0.235 | -7.72 | 0.134 | 2.60 | -0.300 | -8.12 |
| Bicycle owner | -0.029 | -2.38 | -0.033 | -2.08 | -0.054 | -2.05 | 0.069 | 1.53 | -0.061 | -1.84 |
| Log of years living in current location | x | | | | x | | 0.064 | 2.28 | 0.044 | 2.57 |
| Log of home to campus distance | 0.041 | 4.57 | 0.049 | 4.12 | 0.113 | 5.67 | x | | 0.168 | 6.88 |
| Main mode: Bicycle | -0.125 | -4.51 | -0.153 | -4.29 | -0.248 | -4.56 | x | | -0.416 | -6.10 |
| Main mode: Local transit | 0.045 | 2.84 | 0.063 | 3.07 | 0.113 | 3.27 | -0.217 | -3.98 | x | |
| Main mode: Driving alone | 0.076 | 2.80 | 0.066 | 1.78 | 0.167 | 2.78 | x | | 0.236 | 3.15 |
| Main mode: Ride sharing | 0.074 | 2.24 | 0.093 | 2.11 | 0.188 | 2.57 | x | | 0.176 | 1.92 |
| Same modes in Fall and Winter | x | | x | | x | | 0.123 | 1.50 | x | |
| Living in apartment/condo | 0.008 | 0.54 | 0.018 | 0.54 | 0.049 | 1.60 | -0.064 | -1.10 | 0.041 | 1.05 |
| Living on-campus residence | x | | x | | x | | 0.500 | 4.76 | 0.317 | 3.86 |
| Living with partner | 0.039 | 1.79 | 0.030 | 1.01 | 0.071 | 1.48 | 0.132 | 1.68 | 0.205 | 3.85 |
| Living alone | 0.064 | 2.74 | 0.064 | 2.04 | 0.093 | 1.85 | x | | 0.195 | 3.58 |
| Living with roommates | 0.038 | 1.92 | 0.040 | 1.55 | 0.053 | 1.27 | -0.138 | -2.01 | -0.195 | -3.75 |
| Resident phone owner | x | | x | | x | | -0.082 | -1.43 | x | |
| Living in downtown Toronto | -0.037 | -1.53 | -0.050 | -1.80 | -0.093 | -2.07 | x | | 0.195 | 3.58 |
| Home to nearest bus stop | x | | 0.011 | 1.41 | 0.015 | 1.21 | x | | x | |
| Home to nearest rail station | -0.012 | -1.25 | x | | -0.022 | -1.10 | x | | -0.074 | -2.93 |
| Home to nearest subway station | -0.013 | -2.08 | -0.007 | -0.96 | -0.021 | -1.65 | 0.059 | 2.42 | -0.041 | -2.77 |
| Area (sq. km) of 1 km walking buffer | x | | x | | x | | 0.107 | 1.52 | x | |
| Employment density in home TAZ | -0.006 | -1.01 | x | | x | | 0.050 | 2.39 | x | |
| **Dispersion Parameter** | | | 1.562 | 19.18 | | | | | | |
| **Variance of Gamma** | | | | | | | | | | |
| **Heterogeneity** | | | | | 2.195 | 7.58 | | | 1.578 | 6.59 |
| **Ordered Pair Correlation Coefficients: Logit Function** | | | | | | | | | | |
| Population density in home TAZ | -0.170 | -3.20 | -0.105 | -2.17 | | | | | | |
| Age lower than 21 years | 1.419 | 3.88 | 1.393 | 4.78 | | | | | | |
| **Additional Count-Specific Systematic Utility** | | | | | | | | | | |
| Owning a car | | | | | | | | | | |
| 0 telecommuting per week | 0.240 | 3.54 | -0.016 | -0.25 | | | | | | |
| 1 telecommuting per week | -0.084 | -1.31 | -0.192 | -3.25 | | | | | | |
| 2 telecommutings per week | -0.153 | -2.90 | -0.202 | -4.12 | | | | | | |
| Same modes in Fall and Winter | | | | | | | | | | |
| 1 telecommuting per week | -0.500 | -16.15 | -0.219 | -8.40 | | | | | | |
| 2 telecommutings per week | -0.251 | -6.39 | -0.067 | -1.81 | | | | | | |
| Living on campus | | | | | | | | | | |
| 0 telecommuting per week | 0.429 | 5.19 | 0.286 | 4.09 | | | | | | |
| Area of 1 km walking buffer in home zone | | | | | | | | | | |
| 2 telecommutings per week | 0.114 | 4.54 | 0.224 | 7.69 | | | | | | |
| 3 telecommutings per week | -0.113 | -6.43 | 0.105 | 5.18 | | | | | | |
| Threshold parameter for frequency: 0 | | | | | 1.019 | 3.30 | | | x | |
| Threshold parameter for frequency: 1 | | | | | 1.848 | 5.93 | | | 0.507 | 1.40 |
| Threshold parameter for frequency: 2 | | | | | 2.898 | 9.04 | | | 2.047 | 5.47 |
| Threshold parameter for frequency: 3 | | | | | 3.608 | 10.82 | | | 2.937 | 7.44 |
| Threshold parameter for frequency: 4 | | | | | 4.179 | 11.94 | | | 3.643 | 8.61 |
| Threshold parameter for frequency: 5 | | | | | 4.673 | 12.63 | | | 4.261 | 9.28 |
| Threshold parameter for frequency: 6+ | | | | | 5.022 | 12.92 | | | 4.708 | 9.53 |

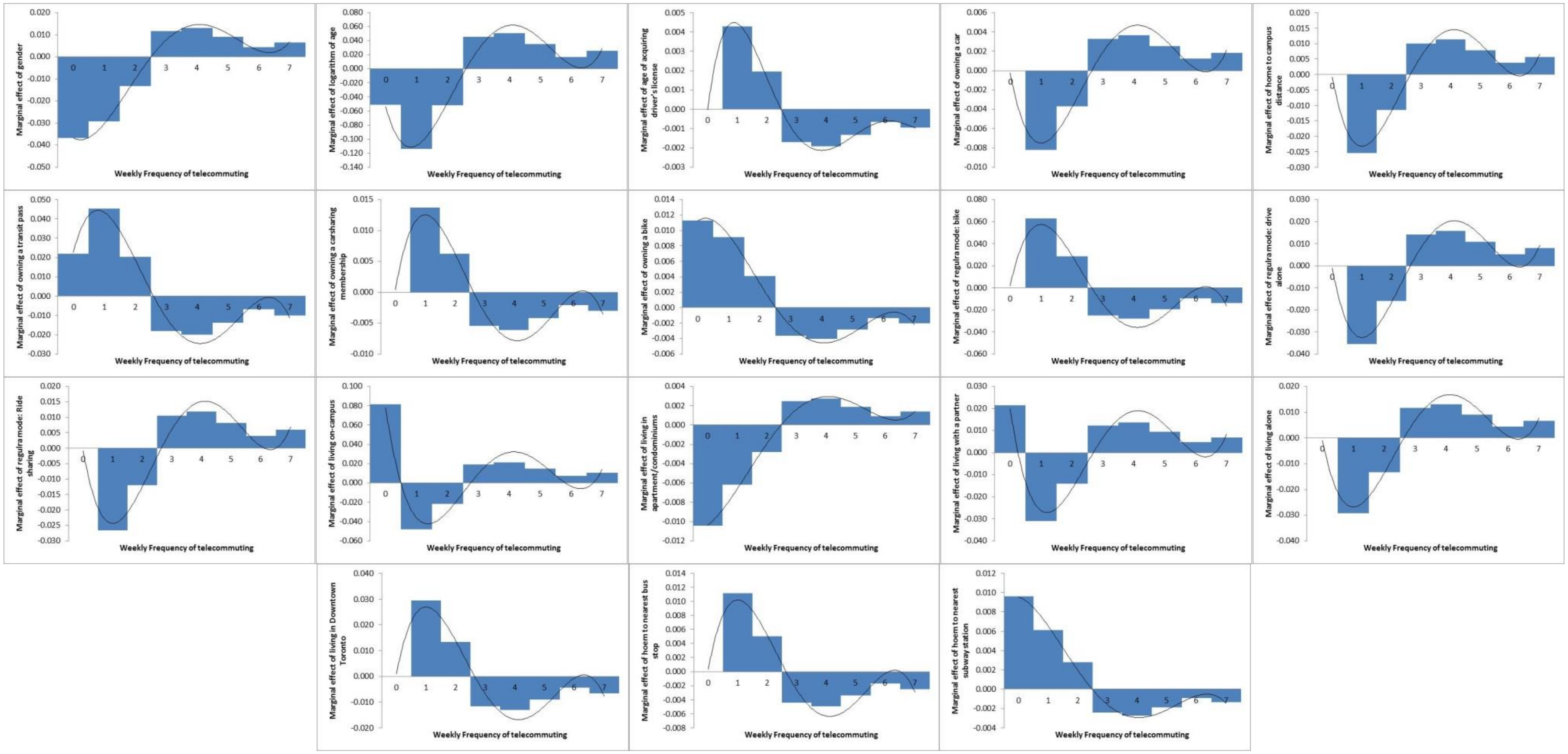

**Figure 2: Marginal effects of covariates in the split population OEV model.**